\documentclass[preprint]{aastex61}         

\usepackage{amsmath}
\usepackage{mathtools}
\usepackage{graphicx}
\usepackage[utf8]{inputenc}
\usepackage{verbatim}

\shorttitle{BRITE photometry of $\beta$~Lyrae in 2018}
\shortauthors{Rucinski et al.}

\begin{document}

\title{
Photometry of $\beta$~Lyrae in 2018 by the BRITE satellites
}

\author{Slavek M.\ Rucinski} 
\affiliation{Department of Astronomy and Astrophysics, 
University of Toronto,\\ \qquad 50 St.~George St., Toronto, Ontario, M5S~3H4, Canada}

\author{Andrzej Pigulski}
\affiliation{Instytut Astronomiczny, Uniwersytet Wroc{\l}awski, 
Kopernika~11, 51-622 Wroc{\l}aw, Poland}

\author{Rainer Kuschnig}
\affiliation{Institute of Communication Networks and Satellite Communications, 
   Graz University of Technology,\\ \qquad Inffeldgasse~12, 8010 Graz, Austria}

\author{Anthony F. J. Moffat} 
\affiliation{D\'{e}partement de physique, Universit\'{e} de Montr\'{e}al, C.P. 6128, 
   Succursale Centre-Ville,\\  \qquad Montr\'{e}al, Qu\'{e}bec, H3C 3J7, Canada}   

\author{Adam Popowicz}
\affiliation{Faculty of Automatic Control, Electronics and Computer Science 
Institute of Automatic Control, 
Politechnika \'{S}laska,\\ \qquad  Akademicka~16, 44-100 Gliwice, Poland}

\author{H.\ Pablo}
\affiliation{American Association of Variable Star Observers, 49~Bay State Road, 
Cambridge, MA~02138, USA}

\author{G.\  A.\  Wade}
\affiliation{Department of Physics and Space Science, 
Royal Military College of Canada,\\  \qquad PO Box~17000 Kingston, Ontario, K7K~7B4, Canada}

\author{Werner W.\ Weiss}
\affiliation{Institut f\"{u}r Astrophysik, Universit\"{a}t Wien, T\"{u}rkenschanzstrasse~17, 1180~Wien, Austria}

\author{Konstanze Zwintz}
\affiliation{Universit\"{a}t Innsbruck, Institut f\"{u}r Astro- und Teilchenphysik, Technikerstra{\ss}e~25, 
A-6020 Innsbruck, Austria}

\correspondingauthor{Slavek M. Rucinski}         
\email{rucinski@astro.utoronto.ca}


\begin{abstract}
Observations of $\beta$~Lyr in four months of 2018 
by three BRITE Constellation satellites (the red-filter BTr and BHr, and 
the blue-filter BLb) permitted a first, limited look into the light-curve variability 
in two spectral bands. The variations were found to be well correlated outside the innermost 
primary minima with the blue variations appearing to have smaller amplitudes than the red;
this reduction may reflect their presumed origin in the cooler, outer parts of
the accretion disk. This result must be confirmed with more extensive  
material as the current conclusions are based on observations spanning
slightly less than three orbital cycles of the binary.  
The assumption of an instrumental problem and the applied corrections
made to explain the unexpectedly large amplitude of the red-filter light-curve 
observed with the BTr satellite in 2016 are fully confirmed by the 2018 results. 
 \end{abstract}


\section{Introduction} 
\label{sec:intro}

The eclipsing binary $\beta$~Lyr (HD~174638, HR~7106; $V_{\rm max}=3.4$ mag, 
$B-V=0.0$ mag) was a target of an extensive, dedicated study of its light-curve variations 
 by the BRITE Constellation satellites by \citet[Paper~I]{Rci2018}. 
 The BRITE Constellation was described by
 \citet{Weiss2014}, \citet{Pablo2016}, and \citet{Pop2017}. 
 The current paper can be considered as a companion and addendum to Paper~I.
 A successful  further re-observation of $\beta$~Lyr is not guaranteed in view of the
 progressive deterioration of the CCD detectors of the red filter
 BRITE satellites (see the text below) so we present here the extant results. 
  
The light-curve variations of $\beta$~Lyr were interpreted in Paper~I in terms 
of unstable accretion of the matter shed by the B6-8~II bright giant ($3 M_\odot$) 
unto an invisible, much more massive companion ($\simeq 13 M_\odot$) in
the interacting binary system with  orbital period of $P = 12.915$ day.
Paper~I utilized a long series of the almost continuous data 
from the satellite ``Toronto'' (BTr), aided in part by the satellite Uni-BRITE (UBr); 
both satellites used red filters.
The observations were done in 2016 and extended over more than 10 revolutions of 
the binary with  uniform, 100-minute sampling. The study led to characterization of the variations 
as a stochastic Gaussian process with the dominant variability time scale of about
$0.65 - 4$ days ($0.05 - 0.3$ in orbital phase), showing slightly stronger correlation 
than the red-noise signal. It was found that the signal decorrelation time scale  
$d = 0.88 \pm 0.23$ days (or $0.068 \pm 0.018$ in phase) follows 
the same accretor-mass dependence 
as that observed for active galactic nuclei and quasi-stellar objects.

While Paper~I was  successful in terms of the temporal characterization 
of the stochastic variability of $\beta$~Lyrae,
it had two deficiencies: (1)~it addressed the light-curve variations in 
only the red-filter spectral band, (2)~an important assumption was made to 
correct the extensive BTr satellite data for a newly discovered instrumental problem. 
This problem was later identified to affect the BRITE red-filter satellites due
to the radiation detector damage. It was detected mostly thanks to the 
large amplitude of the $\beta$~Lyr eclipses which had extended the required 
dynamic range of the CCD response for a single object. The problem found a 
successful explanation in terms of charge-transfer inefficiency (CTI) effects 
\citep{Pig2018,Pop2018} as a loss of a constant fraction of the CCD charge 
leading to a stronger modulation of the remaining part of the signal. 
A  linear transformation was proposed in Paper~I and led to a good 
agreement with simultaneous data from the BTr and UBr satellites. However, the UBr 
observations had a limited time span so that a direct confirmation on the correctness 
of the applied transformation was felt as necessary with the same BTr satellite. 
 
Section~\ref{sec:obs} describes the 2018 observations while Section~\ref{sec:lc}
presents the phased, seasonal light curves needed for determination of 
deviations from them caused by the  variability of $\beta$~Lyrae which are
presented for the red and blue spectral bands in Section~\ref{sec:rb}. Conclusions
in Section~\ref{sec:concl} close the paper.
 
\begin{figure}[ht]
\begin{center}
\includegraphics[width=1.0\textwidth]{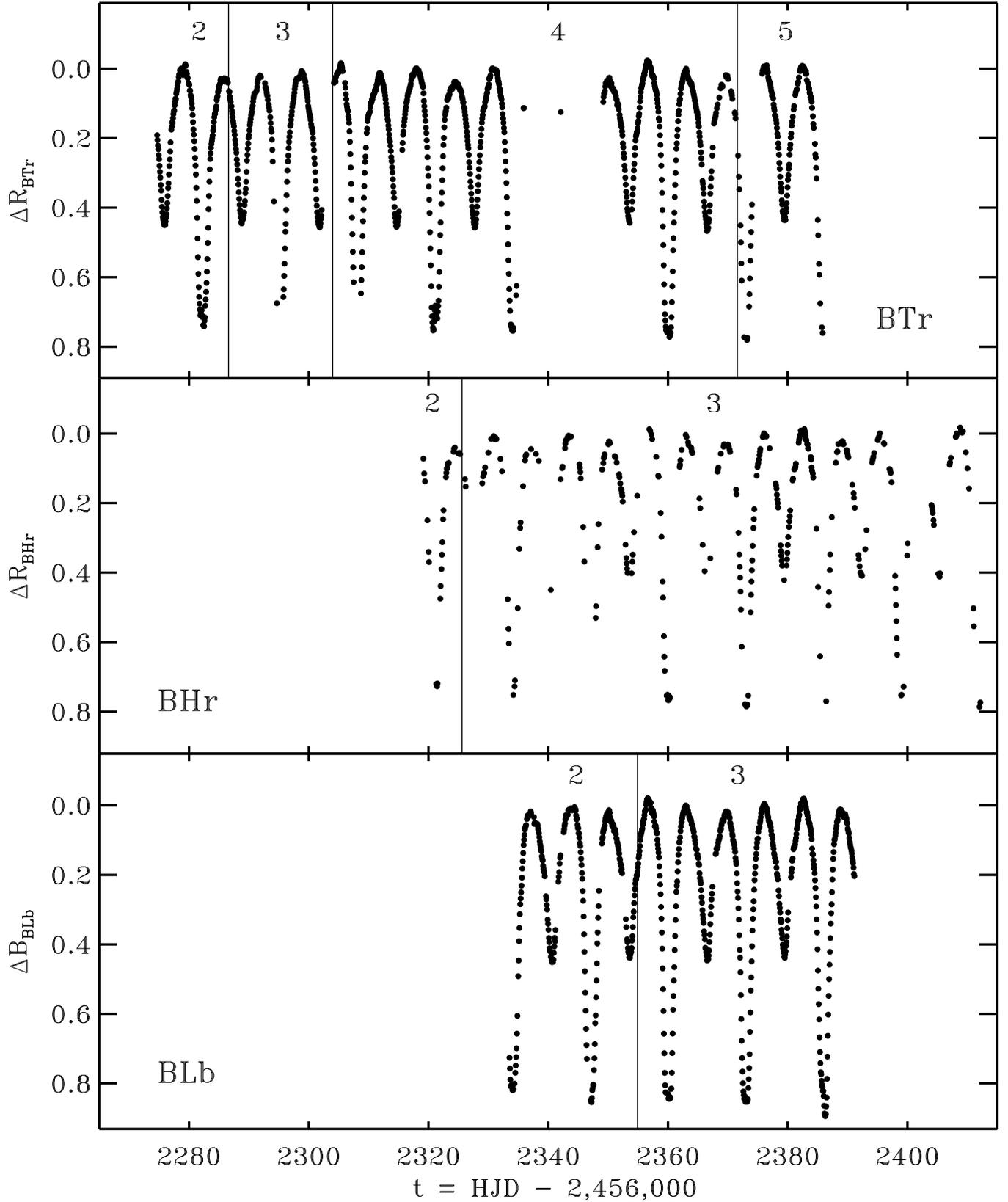}    
\caption{
BRITE observations of $\beta$~Lyrae in 2018 versus
time in $t = HJD- 2\,456\,000$. The satellite-orbit
averages (in magnitudes) shown in the figure 
resulted from standard pipeline and de-correlation processing. 
The magnitudes in the figure have been adjusted to set the zero point 
at light maxima $m_0 = 3.30$ (BTr; the upper panel), 
3.45 (BHr, the middle panel)  and 3.515 for BTr (BLb, the lower panel), 
respectively.
The numbers above the data points label each of the satellite setups (see the text).
}
\label{fig_mags}
\end{center}
\end{figure}

\section{2018 observations} 
\label{sec:obs}

The new observations of $\beta$~Lyrae were obtained between
5 June 2018 and 23 October 2018
using the red-filter satellites BRITE-Toronto (BTr) and BRITE-Heweliusz (BHr), 
and the blue-filter satellite BRITE-Lem (BLb).
The individual exposures (three per minute) 
have been grouped into much better defined satellite-orbit average data, permitting 
brightness sampling at 98.4 min for BTr, 97.7 min for BHr and 99.5 min for BLb. 
The BTr satellite was oriented to place  the $\beta$~Lyr image
in a different CCD location than in 2016 to avoid the CTI-affected area which complicated 
analysis of the otherwise excellent data acquired previously. 

\begin{figure}[ht]
\begin{center}
\includegraphics[width=0.75\textwidth]{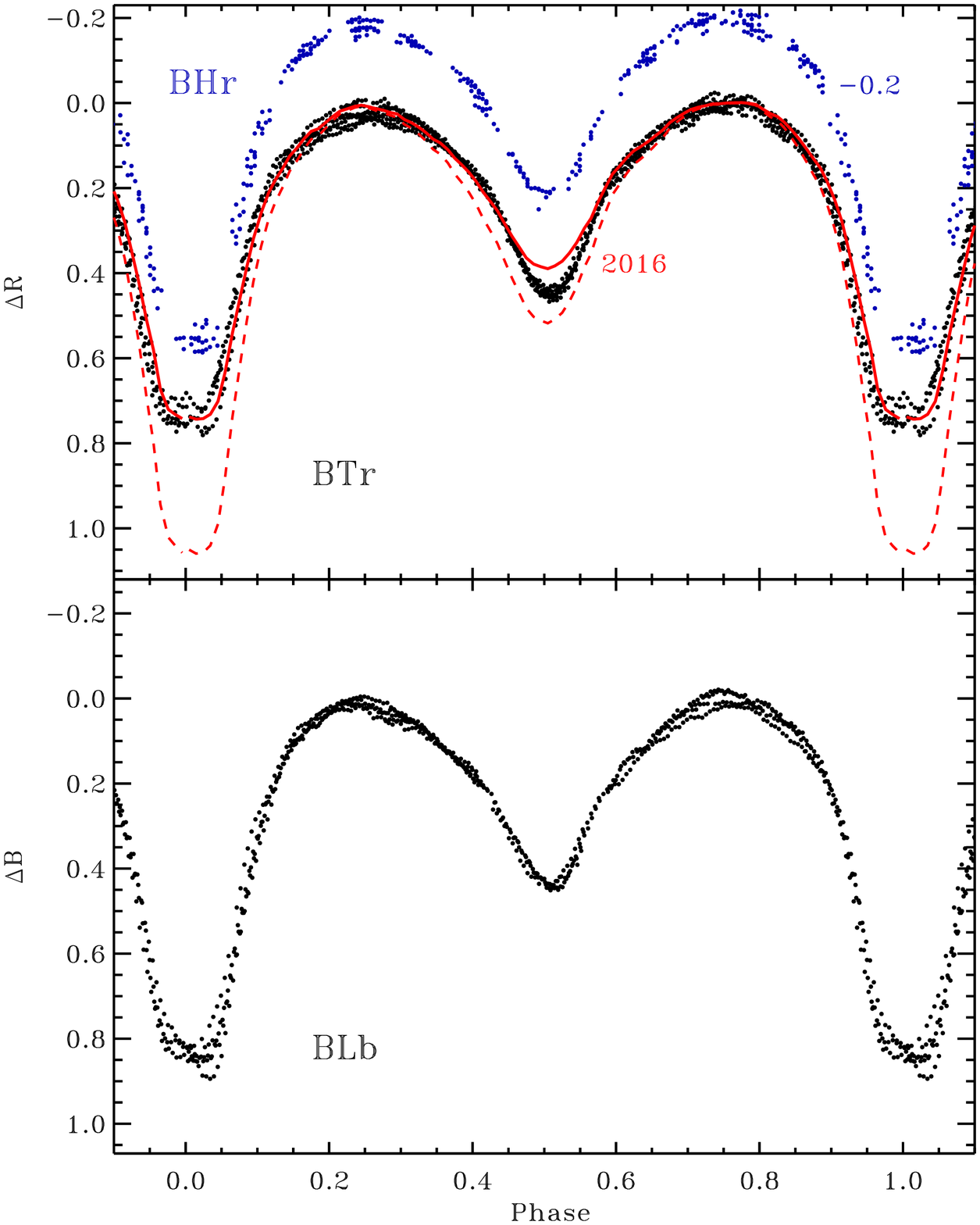}    
\caption{
The data from the three satellites are shown  versus the $\beta$~Lyrae
orbital phase. The results for the red-filter satellites BTr and BHr are
in the upper panel;  the results for the blue-filter satellite BLb are in the lower panel.
The 2018 BTr and BLb observations are marked by black symbols,
while the blue color is used for the 2018 BHr observations (plotted with a shift by 
$-0.2$ mag).
The red lines in the upper panel show the BTr light curves of the 2016 run: 
the continuous line is the assumed, corrected shape of the light curve, as
used in Paper~I, while the broken line shows the mean light curve 
as it was originally observed.  
The magnitude scales were adjusted to the maximum light,
as in Figure~\ref{fig_mags}.
}
\label{fig_lc}
\end{center}
\end{figure}

The 2018 data turned out to be compromised by stabilization problems occurring for 
all the three satellites. This forced the run to be partitioned into several satellite 
setups\footnote{A satellite ``setup'' is a set of positioning instructions for the satellite 
and for the CCD windowing system, as described in the
Appendix of \citet{Pop2017}.}, each with a possibility of
a small shift in the magnitude scale zero-point. The more extensive and denser 
BTr observations were divided  
into four setups  (numbers 2 -- 5), while the BHr and BLb observations
were split into two setups (for each, the numbers 2 -- 3). The initial,
position-acquisition setups (number 1) were discarded.
In terms of the binary orbits, the coverage provided by the satellites
was respectively seven (BTr), seven (BHr) and five (BLb) binary orbits long
with a very uneven distribution of the data points for BHr. 
The data are shown in Figure~\ref{fig_mags} and are listed in 
Table~\ref{tab:avg-sat}.

\placetable{tab:avg-sat}        

\bigskip

\section{The seasonal light curves}
\label{sec:lc}
Although three BRITE satellites were used to observe $\beta$~Lyrae in 2018,
the frequent disruptions (Figure~\ref{fig_mags}) prevent an equally detailed  temporal 
analysis of the light-curve variability as for the 2016 BTr data presented in Paper~I.
Thus, none of the modern statistical tools utilized in Paper~I could be applied 
for the new data although the light-curve variations were definitely present.
In the light curves plotted versus orbital phase
(Figure~\ref{fig_lc}) the eye can separate variations taking place
during individual orbital cycles of the binary. We note that the variations were
smaller during the secondary eclipse confirming the conclusions in Paper~I.
The orbital phases for $\beta$~Lyrae were calculated from 
the same quadratic ephemeris as used in Paper~I, by
\citet{Ak2007},  with the locally linear elements set for $E = 3875$ at
$t_0 = 2347.0119$ and $P_0 = 12.94379$ d. 

The stochastic variability observed in the $\beta$~Lyrae light curve was
characterized in Paper~I by the standard deviation $\sigma = 0.0130$ mag, 
but averaging of several binary cycles and the high 
uniformity of the BTr observations in 2016 resulted
in reduction of the median standard error of the mean, seasonal
light curve to 0.0036 mag per 0.01 phase interval.
The 2018 data are poorer than the 2016 data, mostly because
of the smaller number of data points per phase bin (median numbers
12, 4 and 8 points per 0.01 phase interval for BTr, BHr and BLb, respectively).
While the median errors were increased moderately compared
with the 2016 seasonal light curve, to 
0.0043, 0.0064, and 0.0053 mag, the uncertainties for some of the
phase intervals reached as much as  0.017, 0.021, and 0.021 mag 
(BTr, BHr and BLb, respectively) in the deepest parts of the
eclipses, mostly due to the use of the magnitude
scale which exaggerates errors when the star is relatively
faint -- see the next Section~\ref{sec:rb} where the brightness levels are
discussed in terms of flux units. The BHr seasonal light curve was 
poorly covered with some of the 0.01 phase intervals having only
one or no data points. 
The uncertainties for some phases of the BTr and BLb seasonal
light curves have a dominating impact on the accuracy with which 
variations in the individual light curves could be determined 
(Section~\ref{sec:rb}).
The mean light curves are listed in Table~\ref{tab:mean}.

\placetable{tab:mean}     

Irrespectively of the above qualifications, an encouraging result of the 2018 
observations is a verification of the assumptions in Paper~I,
made to correct the BTr 2016 data for the instrumental CTI problem.
We assumed in Paper~I that 25.5\% of the unmodulated signal  
had been lost, increasing the variability amplitude in the modulated part.
While we did not use the light curves for an eclipsing-model analysis, a 
correction for the CTI problem was needed to  evaluate the scale for the light-curve
variations. The magnitude of the CTI-effect correction was determined in Paper~I
from the parallel UBr satellite observations. 
However, it was felt necessary to  verify 
our assumptions with observations done with the same BTr satellite, 
with the image of $\beta$~Lyrae shifted to an unaffected area of its
CCD. The 2018 BTr data agree with the corrected 2016 light curve, as shown by 
a comparison of the data points and lines in Figure~\ref{fig_lc}; the
BHr mean light curve also fully agrees with that result. 
The amplitude is very well reproduced, including the light-curve sections 
in the deeper minimum, which led us -- for the 2016 data --
to detection of the CTI problem. 
The agreement in the detail of the light-curve shape is somewhat 
less satisfactory, particularly during the secondary minimum, which is otherwise 
best defined thanks to the diminished magnitude of the light-curve 
physical variations in that phase interval.  
The slightly imperfect fit around the secondary minimum 
may indicate that a linear correction for the CTI effect may need some 
adjustment although a nonlinearity would present a considerable challenge 
for the mission in terms of possible ways to diminish its influence on the final
results. 
However, a much simpler solution for this relatively minor discrepancy 
would be to resort to the {\it slow\/} variations in the $\beta$~Lyrae system and the 
well established -- but still unexplained -- 283 day periodicity  
expected to be exactly in anti-phase relative to the 2016 season
with 2.5 cycles separating the two BRITE runs  
(for the list of references relating to the 283-day periodicity,
see Sec.~3 in Paper~I). The cycle was found by \citet{vHW1995}
to have a semi-amplitude of about 0.02 mag in the red part of the spectrum
which could easily accommodate the remaining slight light-curve shape 
discord.

\section{Simultaneous observations in the red and blue spectral bands}
\label{sec:rb}

While the BHr 2018 data were too sparse for a reliable definition 
of the light-curve physical variations (Figure~\ref{fig_mags}), 
the BTr and and BLb observations
offered a possibility of seeing correlation of light-curve 
variations in the red and blue spectral bands. Unfortunately,
the simultaneous successful operations lasted a relatively short
time, $2349 < t < 2386$ (Figure~\ref{fig_mags}), 
i.e.\  about three orbital cycles of the binary. During that time
the blue observations were more uniform in time than the red data 
so that sufficiently simultaneous data from both satellites (within an interval of
one satellite orbit) could be determined for only 394 epochs. 

\begin{figure}[!ht]
\begin{center}
\includegraphics[width=0.50\textwidth]{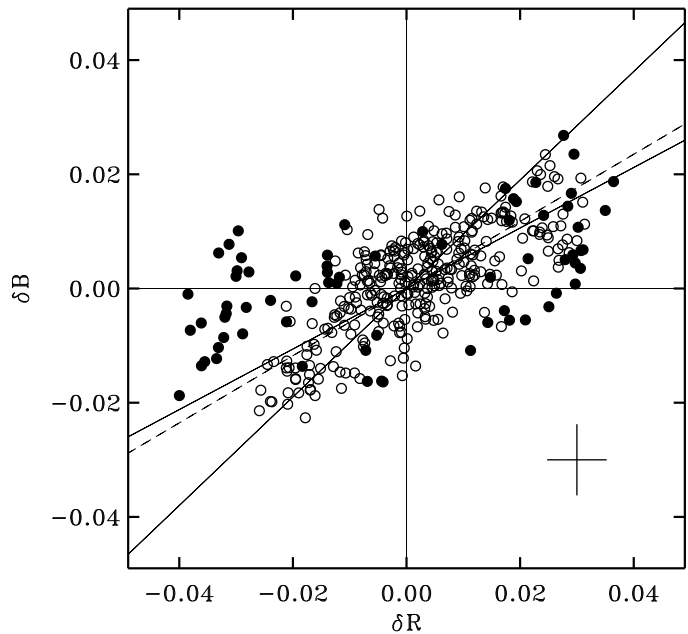}    
\caption{
The correlation diagram of the light-curve deviations $\delta$ 
expressed in flux units with the unit flux assumed at the maximum light
of the binary in the given spectral band. The deviations are for the time
interval $2349 < t < 2386$ when the red- and blue-filter 
satellites BTr and BLb operated sufficiently simultaneously.
The data for the central parts of the primary eclipses within $\pm 0.12$ in 
orbital phase are marked by filled circles. 
Results of the linear fits for the out-of-eclipse phases (open circles) 
are shown by slanted lines:
the solid lines give the least absolute-deviation (LAD) fits for the
red and blue deviations respectively taken as independent variables, while
the broken line gives the least-squares fit (LSQ). 
The error bars in the right lower corner show  the median
uncertainties of $\delta$ in each color.
}
\label{fig_correl}
\end{center}
\end{figure}

The magnitude scale in Figures \ref{fig_mags} 
and \ref{fig_lc} has been useful for pointing out the technical problem 
caused by the CTI defect, but this scale is not appropriate for
consideration of the light-curve variations: It is unphysical and nonlinear, and
-- when the magnitudes are related to the maximum light of the binary -- it
emphasizes the variations during the low-brightness light-curve sections dominated 
by geometrical eclipse effects. For analysis of the light-curve variations and
for consistency with Paper~I, the light curves for BTr and BLb were
transformed into fluxes and then used to find the light-curve flux deviations
$\delta$ as the differences between the
individual flux and the mean, seasonal light curve, following the definition in Paper~I. 
The maximum light of the binary was estimated (to $\pm 0.005$ mag) 
for use as the flux unit in the respective spectral ranges
at $m_0 = 3.30$ and 3.515 for the BTr and BLb satellites. The      
deviations  are listed in Table~\ref{tab:avg-sat}.  
The individual uncertainties of $\delta$ were estimated as quadratically
added uncertainties of the mean light curves (increased by $\sqrt{2}$ to
account for interpolation and smoothing) 
and the uncertainties of the satellite-orbit flux averages. 
The resulting uncertainties are dominated by those of the seasonal
light curves with median standard errors of 0.0051 and 0.0064     
(in flux units) for the BTr and BLb satellites. The maximum errors,
particularly at the slopes of the eclipses  reached 
0.0119 and 0.0130 for the two satellites, respectively.                   

\placetable{tab:devs}     

The deviations $\delta$ with their estimated uncertainties 
are listed in Table~\ref{tab:devs} and are shown in 
the two-color correlation diagram in Figure~\ref{fig_correl}. 
Since light-curve variations appeared to be 
larger within the innermost primary eclipses,  
the deviations within the phase interval $\pm 0.12$ 
are marked in Figure~\ref{fig_correl} as filled circles. 
The red-color variations appeared to be very large 
and poorly correlated with the blue variations 
in the deepest parts of the primary eclipses. However,
we note that the seasonal light curves in both filters were particularly
poorly defined exactly within the phases of the deepest primary eclipse. 

The two color variations at the phases outside of the interval $\pm 0.12$, 
324 in number, correlate moderately well with the correlation coefficients:
Spearman $c_\rho = 0.70$, Kendall (rank): $c_\tau = 0.51$. 
A least-squares (LSQ) fit to the deviations assuming the errors to 
affect both coordinates, with weights calculated from the
error estimates as described above, 
gave: $\delta \! B = (0.0004\pm0.0004) + (0.587\pm0.034) \, \delta \! R$.
This result is  confirmed by the bootstrap technique (repeated multiple LSQ
solutions with random data sampling):
$\delta \! B = 0.00035^{+0.00035}_{-0.00032} + 0.529^{+0.030}_{-0.029} \, \delta \! R$.
It is well recognized that the LSQ is sensitive to outliers; a more
robust estimate can be provided by the least-absolute-deviation (LAD)
technique. The LAD does not provide algorithmic estimates 
of the parameter errors, so that 
a standard approach to handle the outliers in both coordinates is mutual inversion 
of coordinates as  independent variable. The LAD fits are:
$\delta \! B = 0.0004 + 0.530 \, \delta \! R$ for the direct solution
 $\delta \! B$ versus $\delta \! R$ (orientation as in Figure~\ref{fig_correl}),  
and $\delta B = 0.0001 + 0.950 \, \delta \! R$ from the inverted solution
($\delta \! B$ as the independent variable),
after the back-transformation of the variables. 
The three fits are shown in Figure~\ref{fig_correl} as lines with the LAD 
results bracketing the LSQ fit. We believe that
the spread of the two LAD solutions is more representative of the level of uncertainty 
in this problem than the LSQ errors so that  slope close
to one is not fully excluded, although the formal result -- which definitely requires 
observational confirmation -- is that the blue filter light-curve variations are smaller 
than those in the red. 
Although this result may be considered as unexpected in view of the blue color
of the star, we should note that \citet{hp1991} in their careful analysis 
of spectrophotometric properties of the complex spectrum of $\beta$~Lyr 
assigned a cooler spectral component -- identified as having A8~II characteristics --
to be produced at the rim of the accretion disk.
Thus, any instabilities produced by the ongoing accretion and taking place 
at or within the rim may be expected to be redder than the average color 
of the binary system which is dominated by the B6-8~II bright giant. If our finding
is repeated by more extensive observations, this will be a confirmation
that by studying the light-curve variations we almost directly see
the accretion phenomena in $\beta$~Lyr.

\section{Conclusions}
\label{sec:concl}
The 2018 data for $\beta$~Lyr were less uniform than the red-filter data
analyzed in Paper~I and thus did not permit an equally 
extensive variability analysis as in that paper. 
But they offered a first look -- unfortunately only during three orbital cycles of
the binary -- into the mutual relation of the two-color light-curve variations 
in the dominant range of the temporal scales of 0.65 -- 4 days.
Most interestingly, the blue-band variations appear to have smaller 
amplitudes than the red variations, although this result definitely needs 
confirmation.
In 2016, the blue-filter satellite BLb was not functional at that time
so that we had no choice but to base the analysis
for Paper~I solely on the red-filter data from the BTr satellite. These data 
were extensive and uniform, but suffered from an unwelcome and initially 
puzzling amplitude problem, later explained by a CTI instrumental effect. Yet --
if the larger amplitude of the variation is confirmed 
 -- these data have been fortuitously obtained in the better-suited of the two
available spectral ranges.
The result of the diminished magnitude of the blue
light curve variations compared with the red ones is new and definitely 
must be verified observationally by more extensive observations.
In terms of a possible physical explanation, the effect 
is in agreement with the rim of the $\beta$~Lyr secondary-component 
accretion disk as a location for the light-curve variability. 

The 2018 data for the red-filter satellite BTr fully confirm the validity of the
assumed corrections for the instrumental CTI effect as applied in Paper~I
to the 2016 data. 

\begin{acknowledgements}
The study is based on data collected by the BRITE Constellation satellite mission, 
designed, built, launched, operated and supported by the Austrian Research 
Promotion Agency (FFG), the University of Vienna, the Technical University of Graz, 
the University of Innsbruck, the Canadian Space Agency (CSA), the University of 
Toronto Institute for Aerospace Studies (UTIAS), the Foundation for Polish 
Science \& Technology (FNiTP MNiSW), and Polish National Science Centre (NCN).
The operation of the Polish BRITE satellites is funded by a SPUB grant from the 
Polish Ministry of Science and Higher Education (MNiSW). 

The research of SMR and AFJM has been supported by the
Natural Sciences and Engineering Research Council (NSERC) of Canada. 
APi acknowledges support from the National Science Centre (NCN) 
grant 2016/21/B/ST9/01126.
APo was responsible for image processing and automation of photometric 
routines for the BRITE Constellation with support from the statutory-activities 
grant BK/200/RAU1/2018 t.3.
RK, WW and KZ acknowledge support by the Austrian Space Application 
Programme (ASAP) of the Austrian Research Promotion Agency (FFG).
\end{acknowledgements}


---------------------------------------------------------------------------------

\begin{deluxetable}{cccccc}

\footnotesize
\tablecaption{The satellite-orbit, average data 
\label{tab:avg-sat}
}
\tablenum{1}

\tablehead{\colhead{Sat} & \colhead{$t$}  &  \colhead{$\phi$} & \colhead{$m$} & 
      \colhead{$\sigma(m)$}  &\colhead{$N$} \\
      \colhead{(1)} & \colhead{(2)}  &  \colhead{(3)} & \colhead{(4)} & 
      \colhead{(5)}  &\colhead{(6)}        
}

\startdata
 1 & 2274.6697  &  0.4110  &  3.4914 &  0.0016 &  29 \\
 1 & 2274.7375  &  0.4163  & 3.5055  &  0.0009 &  26 \\
 1 & 2274.8057  &  0.4216  & 3.5209  &  0.0015 &  29 \\
 1 & 2274.8740  &  0.4268  & 3.5317  &  0.0021 &  30 \\
 1 & 2274.9424  &  0.4321  & 3.5485  &  0.0017 &  28 \\
 \enddata

\tablecomments{
The columns: 
(1)~The satellite code (1 = BTr, 2 = BHr, 3 = BLb).
(2)~Time $t$: The mean heliocentric time $t = HJD- 2\,456\,000$.
(3)~ $\phi$: The binary orbital phase calculated
using the locally linear elements based on the quadratic ephemeris by 
\citet{Ak2007} for $E = 3875$: $t_0 = 2347.0119$ and $P_0 = 12.94379$ d.
(4)~$m$: The magnitude with an arbitrary zero point from the
BRITE pipeline processing. 
(5)~$\sigma(m)$: The error of $m$ from the scatter of individual
$N$ observations.
(6)~ $N$: The number of averaged observations.\\
The data consist of the satellite setups partitioned at: \\
BTr: $t =  2286.6, 2304.6, 2371.6$; \\
BHr: $t = 2325.6$; \\
BLb: $t = 2354.9$.
\\
The table is published in its entirety in machine-readable format.
A portion is shown here for guidance regarding its form and content.}

\end{deluxetable}


\begin{deluxetable}{ccccc}

\footnotesize
\tablecaption{The mean, seasonal, 2018 light curves 
\label{tab:mean}}
\tablenum{2}

\tablehead{\colhead{Sat} & \colhead{$\phi$} & \colhead{$m$} & 
      \colhead{$\sigma(m)$}  &\colhead{$N$} 
}

\startdata
    1  & 0.0041  &  4.0211  &  0.0120  &  6\\
    1  & 0.0147  &  4.0335  &  0.0096  &  5\\
    1  & 0.0252  &  4.0462  &  0.0102  &  8\\
    1  & 0.0359  &  4.0018  &  0.0167  &  7\\
    1  & 0.0457  &  3.9527  &  0.0127  &  11\\
 \enddata

\tablecomments{
(1)~The satellite code (1 = BTr, 2 = BHr, 3 = BLb).
(2)~ $\phi$: The average orbital phase of the binary in intervals 
of 0.01 in phase. The phases are calculated
using the linear elements as given in Table~\ref{tab:avg-sat}.
(3)~$m$: The mean magnitude with an arbitrary zero point from the
BRITE pipeline processing. 
(4)~$\sigma(m)$: The error of $mag$ from the scatter of points in 
phase interval of 0.01. 
(5)~ $N$: The number of averaged observations.\\
\\
The table is published in its entirety in machine-readable format.
A portion is shown here for guidance regarding its form and content.}

\end{deluxetable}


\begin{deluxetable}{lccccc}      

\footnotesize
\tablecaption{The light-curve flux deviations in two colors 
\label{tab:devs}}
\tablenum{3}

\tablehead{\colhead{$t$}  &  \colhead{$\phi$} & \colhead{$\delta\!R$} & \colhead{$\varepsilon\!R$} &
      \colhead{$\delta\!B$}  &  \colhead{$\varepsilon\!B$}
}

\startdata   
   2349.1575  &  0.1658  &  $-0.0034$         &  0.0073  &  $-0.0017$   &  0.0063 \\
   2349.2271  &  0.1711  &  $\phd 0.0032$  &  0.0070  &  $-0.0035$   &  0.0059 \\
   2349.2961  &  0.1765  &  $\phd 0.0055$  &  0.0070  &  $-0.0054$   &  0.0047 \\
   2349.3652  &  0.1818  &  $-0.0006$         &  0.0070  &  $-0.0122$   &  0.0047 \\
   2349.4341  &  0.1871  &  $-0.0004$         &  0.0071  &  $-0.0104$   &  0.0045 \\
 \enddata

\tablecomments{
(1)~Time $t$: Mean heliocentric time $t = HJD- 2\,456\,000$.
(2)~ $\phi$: The $\beta$~Lyr orbital phase calculated
for the linear elements as in Table~\ref{tab:avg-sat}.
(3)--(4)~$\delta \! R$ and $\varepsilon \! R$: Deviation of the observed value of the flux
from the seasonal light curve in the red color and its estimated uncertainty as
observed by the BTr satellite.
(4)~$\delta \! B$  and $\varepsilon \! B$: Deviation of the observed value of the flux
from the seasonal light curve in the blue color and its estimated uncertainty as
observed by the BLb satellite.\\  
The fluxes have been normalized to unity at the maximum brightness of
the binary system for each spectral band assuming $m_0 = 3.30$ and
3.515 for the BTr and BLb satellites, respectively.\\
The table is published in its entirety in machine-readable format.
A portion is shown here for guidance regarding its form and content.}
\end{deluxetable}

\end{document}